# Optimizing Quasi-Orthogonal STBC Through Group-Constrained Linear Transformation

*Abstract* — In this paper, we first derive the generic algebraic structure of a Quasi-Orthogonal STBC (QO-STBC). Next we propose *Group-Constrained Linear Transformation* (GCLT) as a means to optimize the diversity and coding gains of a QO-STBC with square or rectangular QAM constellations. Compared with QO-STBC with constellation rotation (CR), we show that QO-STBC with GCLT requires only half the number of symbols for joint detection, hence lower maximum-likelihood decoding complexity. We also derive analytically the optimum *GCLT parameters* for QO-STBC with square QAM constellation. The optimized QO-STBCs with GCLT are able to achieve full transmit diversity, and have negligible performance loss compared with QO-STBCs with CR at the same code rate.

*Index Terms* — Constellation Rotation, Group-Constrained Linear Transformation, Quasi-Orthogonality Constraint, Quasi-Orthogonal Space-Time Block Code.

## I. INTRODUCTION

Orthogonal Space-Time Block Code (O-STBC) offers full transmit diversity with linear decoding complexity [1]. Unfortunately, O-STBC suffers from reduced code rate when complex constellations are necessitated by high transmission rate requirement, and when the required transmit diversity is greater than two. As a result, Quasi-Orthogonal STBCs (QO-STBC) were proposed. Some well known examples include the QO-STBC from [2], the ABBA code from [3] and the transmit diversity scheme from [4,5]. With its quasi-orthogonal code structure, the maximum-likelihood (ML) decoding of a QO-STBC can be performed by searching over (or joint detection of) only a subset of the total number of transmitted symbols, hence the decoding complexity of *quasi-orthogonal* STBC is lower than the general *non-orthogonal* STBC.

The first-generation QO-STBCs, however, could not achieve full transmit diversity. Fortunately, this problem was solved by the technique of Constellation Rotation (CR) [6-10]. To date, full-rate full-diversity QO-STBC for four transmit antennas can be ML-decoded by the joint detection of at least two complex symbols [6-10]. For eight transmit antennas, full-diversity QO-STBC requires joint detection of at least two complex symbols at a code rate of 3/4 [8], or four complex symbols at a code rate of 1 [5,10].

In this paper, we shall show that the number of symbols required for the joint detection of the existing full-diversity QO-STBCs *with square or rectangular regular QAM constellations* can actually be halved if, instead of CR, a novel "*Group-Constrained Linear Transformation* (GCLT)" is used to optimize the original QO-STBCs. To explain the principles of the proposed GCLT, the generic algebraic structure of QO-STBC is first derived in this paper. We then examine the algebraic structure of existing QO-STBCs and study the impact of CR on their decoding complexity. Next, we derive analytically the optimal *GCLT parameters* for a full-rate QO-STBC for four transmit antennas [2] and a rate-3/4 QO-STBC for eight transmit antennas [2] with square QAM constellation. While the optimum GCLT parameters for a full-rate QO-STBC for eight transmit antennas [3] is obtained by computer search. The bit error rate (BER) performance of QO-STBC designed using CR and GCLT are then compared.

## II. QO-STBC AND ITS SIGNAL MODEL

### A. Signal Model for QO-STBC with QAM Constellation

Suppose that there are $N_t$ transmit antennas, $N_r$ receive antennas, and an interval of $T$ symbols during which the propagation channel is constant and known to the receiver. The transmitted signal can be written as a $T \times N_t$ matrix **C** that governs the transmission over the $N_t$ antennas during the $T$ symbol intervals. It is assumed that the data sequence has been broken into blocks with $K$ square or rectangular regular QAM symbols, $x_1, x_2, \ldots, x_K$, in each block for transmission over $T$ symbol periods of time. The code rate of a QO-STBC is defined as $R = K/T$. If square or rectangular regular QAM constellation is used, every complex

QAM symbol can be treated as two independent real PAM symbols. With this and the modeling approach in [11], a STBC **C** can be expressed as:

$$\mathbf{C} = \sum_{q=1}^{K}(s_q\mathbf{A}_q + s_{K+q}\mathbf{A}_{K+q}) = \sum_{p=1}^{2K}(s_p\mathbf{A}_p) \qquad (1)$$

where the transmitted symbols $x_q = s_q + js_{K+q}$ for $1 \leq q \leq K$. The matrices $\mathbf{A}_p$ of size $T \times N_t$, for $1 \leq p \leq 2K$, are called the "dispersion matrices". To limit the total transmission power, they must conform to the power distribution constraint [11]:

$$\text{tr}(\mathbf{A}_p^H \mathbf{A}_p) = TN_t / K \qquad (2)$$

The received signal model can be modeled as [11]:

$$\tilde{\mathbf{r}} = \sqrt{\rho/N_t}\,\mathbf{H}\tilde{\mathbf{s}} + \tilde{\boldsymbol{\eta}} \qquad (3)$$

where the normalization factor $\sqrt{\rho/N_t}$ is to ensure that the SNR ($\rho$) at the receiver is independent of the number of transmit antennas, and

$$\tilde{\mathbf{r}} = \begin{bmatrix} \mathbf{r}_1^R \\ \mathbf{r}_1^I \\ \vdots \\ \mathbf{r}_{N_r}^R \\ \mathbf{r}_{N_r}^I \end{bmatrix},\; \tilde{\mathbf{s}} = \begin{bmatrix} s_1 \\ \vdots \\ s_K \\ \vdots \\ s_{2K} \end{bmatrix},\; \tilde{\boldsymbol{\eta}} = \begin{bmatrix} \boldsymbol{\eta}_1^R \\ \boldsymbol{\eta}_1^I \\ \vdots \\ \boldsymbol{\eta}_{N_r}^R \\ \boldsymbol{\eta}_{N_r}^I \end{bmatrix},\; \mathbf{H} = \begin{bmatrix} \mathcal{A}_1\underline{\mathbf{h}}_1 & \cdots & \mathcal{A}_K\underline{\mathbf{h}}_1 & \cdots & \mathcal{A}_{2K}\underline{\mathbf{h}}_1 \\ \vdots & \ddots & \vdots & \ddots & \vdots \\ \mathcal{A}_1\underline{\mathbf{h}}_{N_r} & \cdots & \mathcal{A}_K\underline{\mathbf{h}}_{N_r} & \cdots & \mathcal{A}_{2K}\underline{\mathbf{h}}_{N_r} \end{bmatrix},\; \mathcal{A}_p = \begin{bmatrix} \mathbf{A}_p^R & -\mathbf{A}_p^I \\ \mathbf{A}_p^I & \mathbf{A}_p^R \end{bmatrix},\; \underline{\mathbf{h}}_i = \begin{bmatrix} \mathbf{h}_i^R \\ \mathbf{h}_i^I \end{bmatrix}.$$

In the above equations, the superscript R and I denote the real part and imaginary part of a scalar, vector or matrix respectively. The $\mathbf{r}_i$ and $\boldsymbol{\eta}_i$, for $1 \leq i \leq N_r$, are $T \times 1$ column vectors which contain the received signal and the zero-mean unit-variance AWGN noise samples for the $i^{th}$ receive antenna over $T$ symbol periods respectively, $\mathbf{h}_i$ is a $N_t \times 1$ column vector that contains $N_t$ independent Rayleigh flat fading coefficients between the $j^{th}$ transmit antenna and the $i^{th}$ receive antenna, $h_{j,i}$, for $1 \leq j \leq N_t$.

*B. Review of QO-STBC with Constellation Rotation*

In this paper, the rate-1 QO-STBC in [2] for four transmit antennas (herein called the code Q4), the 3/4-rate QO-STBC in [2] for eight transmit antennas (herein called the code Q8) and the rate-1 QO-STBC in [3] for eight transmit antennas (herein called the code T8) will be used as representative code examples. First, the code matrix of the Q4 code, $\mathbf{C_{Q4}}$, is shown in (4):

$$\mathbf{C_{Q4}} = \begin{bmatrix} x_1 & x_2 & x_3 & x_4 \\ -x_2^* & x_1^* & -x_4^* & x_3^* \\ -x_3^* & -x_4^* & x_1^* & x_2^* \\ x_4 & -x_3 & -x_2 & x_1 \end{bmatrix} \tag{4}$$

After appropriate CR, Q4 can achieve full diversity with joint detection of two complex symbols for ML decoding [8,10]. The resultant code, called Q4_CR in this paper, has code matrix $\mathbf{C_{Q4\_CR}}$ as shown in (5).

$$\mathbf{C_{Q4\_CR}} = \begin{bmatrix} x_1 & x_2 & x_3 e^{j\pi/4} & x_4 e^{j\pi/4} \\ -x_2^* & x_1^* & -(x_4 e^{j\pi/4})^* & (x_3 e^{j\pi/4})^* \\ -(x_3 e^{j\pi/4})^* & -(x_4 e^{j\pi/4})^* & x_1^* & x_2^* \\ x_4 e^{j\pi/4} & -x_3 e^{j\pi/4} & -x_2 & x_1 \end{bmatrix} \tag{5}$$

where the factor $e^{j\pi/4}$ denotes the CR angle for the QAM symbols $x_3$ and $x_4$.

The ML decoding metrics for Q4_CR is shown in (6). It is derived based on the ML decoding metrics of Q4 from [2], but taking CR into account. We can see that the decoding decision for symbols $x_1$ and $x_4$ is obtained by minimizing the metric $f_{14}$, similarly the decoding decision for symbols $x_2$ and $x_3$ is obtained by minimizing the metric $f_{23}$. Clearly, decoding of $x_1$ and $x_4$ can be performed separately from the decoding of $x_2$ and $x_3$. Since $x_1$ and $x_4$ (or $x_2$ and $x_3$) are each a complex symbol, their ML decoding requires the joint detection of two *complex* symbols (i.e. four real symbols) in total.

$$f_{14}(x_1, x_4) = \sum_{r=1}^{N_r} \left[ (\sum_{n=1}^{4} |h_{n,r}|^2)(|x_1|^2 + |x_4|^2) + 2\operatorname{Re} \left\{ \begin{array}{l} x_1(-h_{1,r}r_1^* - h_{2,r}^*r_2 - h_{3,r}^*r_3 - h_{4,r}r_4^*) + \\ x_4 e^{j\pi/4}(-h_{4,r}r_1^* + h_{3,r}^*r_2 + h_{2,r}^*r_3 - h_{1,r}r_4^*) + \\ x_1 x_4^* e^{-j\pi/4} \times 2\operatorname{Re}(h_{1,r}h_{4,r}^* - h_{2,r}h_{3,r}^*) \end{array} \right\} \right]$$

$$f_{23}(x_2, x_3) = \sum_{r=1}^{N_r} \left[ (\sum_{n=1}^{4} |h_{n,r}|^2)(|x_2|^2 + |x_3|^2) + 2\operatorname{Re} \left\{ \begin{array}{l} x_2(-h_{2,r}r_1^* + h_{1,r}^*r_2 - h_{4,r}^*r_3 + h_{3,r}r_4^*) + \\ x_3 e^{j\pi/4}(-h_{3,r}r_1^* - h_{4,r}^*r_2 + h_{1,r}^*r_3 + h_{2,r}r_4^*) + \\ x_2 x_3^* e^{-j\pi/4} \times 2\operatorname{Re}(-h_{1,r}h_{4,r}^* + h_{2,r}h_{3,r}^*) \end{array} \right\} \right]$$

(6)

where $x_1$ to $x_4$ in (6) are each non-rotated *complex* constellation symbols.

We shall show in Section IV that by optimizing Q4 with the proposed GCLT instead of CR, the resultant code can be decoded with joint detection of only two *real* symbols, while still achieving full transmit diversity gain and full code rate.

## III. QUASI-ORTHOGONALITY CONSTRAINT

### A. *Algebraic Structure of QO-STBC*

In order to quantify the number of symbols required for joint detection, we now derive the algebraic structure of generic QO-STBC, called the Quasi-Orthogonality (QO) Constraint. The concept of QO-STBC is to divide the $K$ transmitted symbols of a codeword into $G$ independent groups, such that symbols in any group are orthogonal to all symbols in the other groups after appropriate matched filtering, while strict orthogonality among the symbols within a group is not required. As a result, the received symbols can be separated into $G$ independent groups by simple linear processing, such that the ML decoding of different groups can be performed separately and in parallel, and the ML decoding of every group can be achieved by jointly detecting only $K/G$ complex symbols that are within the same group.

*Definition 1*: A quasi-orthogonal design $\mathbf{C} = \sum_{p=1}^{2K}(s_p \mathbf{A}_p)$ is such that, when multiplied with the channel fading coefficients to obtain $\mathbf{H}$ as defined in (3), $\mathbf{H}^T\mathbf{H}$ is block-diagonal and consists of $G$ smaller sub-matrices each with size $(2K/G) \times (2K/G)$.

To derive the QO-Constraint, let us multiply a matched filter ($\mathbf{H}^T$) to the received signal $\tilde{\mathbf{r}}$ in (3), and consider a snapshot of $\mathbf{H}^T\mathbf{H}$ as shown below:

$$\mathbf{H}^T\mathbf{H} = \begin{bmatrix} \vdots & \cdots & \vdots \\ \underline{\mathbf{h}}_1^T \mathcal{A}_p^T & \cdots & \underline{\mathbf{h}}_{N_r}^T \mathcal{A}_p^T \\ \underline{\mathbf{h}}_1^T \mathcal{A}_u^T & \cdots & \underline{\mathbf{h}}_{N_r}^T \mathcal{A}_u^T \\ \underline{\mathbf{h}}_1^T \mathcal{A}_q^T & \cdots & \underline{\mathbf{h}}_{N_r}^T \mathcal{A}_q^T \\ \underline{\mathbf{h}}_1^T \mathcal{A}_v^T & \cdots & \underline{\mathbf{h}}_{N_r}^T \mathcal{A}_v^T \\ \vdots & \cdots & \vdots \end{bmatrix} \begin{bmatrix} \cdots & \mathcal{A}_p \underline{\mathbf{h}}_1 & \mathcal{A}_u \underline{\mathbf{h}}_1 & \mathcal{A}_q \underline{\mathbf{h}}_1 & \mathcal{A}_v \underline{\mathbf{h}}_1 & \cdots \\ & \vdots & \vdots & \vdots & \vdots & \\ \cdots & \mathcal{A}_p \underline{\mathbf{h}}_{N_r} & \mathcal{A}_u \underline{\mathbf{h}}_{N_r} & \mathcal{A}_q \underline{\mathbf{h}}_{N_r} & \mathcal{A}_v \underline{\mathbf{h}}_{N_r} & \cdots \end{bmatrix}$$

$$= \begin{bmatrix} \cdots & \cdots & \cdots & \cdots & \cdots & \cdots \\ \vdots & \sum_{i=1}^{N_R} \underline{\mathbf{h}}_i^T(\mathcal{A}_p^T \mathcal{A}_p)\underline{\mathbf{h}}_i & \sum_{i=1}^{N_R} \underline{\mathbf{h}}_i^T(\mathcal{A}_p^T \mathcal{A}_u)\underline{\mathbf{h}}_i & \boxed{\sum_{i=1}^{N_R} \underline{\mathbf{h}}_i^T(\mathcal{A}_p^T \mathcal{A}_q)\underline{\mathbf{h}}_i} & \sum_{i=1}^{N_R} \underline{\mathbf{h}}_i^T(\mathcal{A}_p^T \mathcal{A}_v)\underline{\mathbf{h}}_i & \vdots \\ \vdots & \sum_{i=1}^{N_R} \underline{\mathbf{h}}_i^T(\mathcal{A}_u^T \mathcal{A}_p)\underline{\mathbf{h}}_i & \sum_{i=1}^{N_R} \underline{\mathbf{h}}_i^T(\mathcal{A}_u^T \mathcal{A}_u)\underline{\mathbf{h}}_i & \sum_{i=1}^{N_R} \underline{\mathbf{h}}_i^T(\mathcal{A}_u^T \mathcal{A}_q)\underline{\mathbf{h}}_i & \sum_{i=1}^{N_R} \underline{\mathbf{h}}_i^T(\mathcal{A}_u^T \mathcal{A}_v)\underline{\mathbf{h}}_i & \vdots \\ \vdots & \boxed{\sum_{i=1}^{N_R} \underline{\mathbf{h}}_i^T(\mathcal{A}_q^T \mathcal{A}_p)\underline{\mathbf{h}}_i} & \sum_{i=1}^{N_R} \underline{\mathbf{h}}_i^T(\mathcal{A}_q^T \mathcal{A}_u)\underline{\mathbf{h}}_i & \sum_{i=1}^{N_R} \underline{\mathbf{h}}_i^T(\mathcal{A}_q^T \mathcal{A}_q)\underline{\mathbf{h}}_i & \sum_{i=1}^{N_R} \underline{\mathbf{h}}_i^T(\mathcal{A}_q^T \mathcal{A}_v)\underline{\mathbf{h}}_i & \vdots \\ \vdots & \sum_{i=1}^{N_R} \underline{\mathbf{h}}_i^T(\mathcal{A}_v^T \mathcal{A}_p)\underline{\mathbf{h}}_i & \sum_{i=1}^{N_R} \underline{\mathbf{h}}_i^T(\mathcal{A}_v^T \mathcal{A}_u)\underline{\mathbf{h}}_i & \sum_{i=1}^{N_R} \underline{\mathbf{h}}_i^T(\mathcal{A}_v^T \mathcal{A}_q)\underline{\mathbf{h}}_i & \sum_{i=1}^{N_R} \underline{\mathbf{h}}_i^T(\mathcal{A}_v^T \mathcal{A}_v)\underline{\mathbf{h}}_i & \vdots \\ \cdots & \cdots & \cdots & \cdots & \cdots & \cdots \end{bmatrix} \quad (7)$$

where $1 \leq p, q, u, v \leq 2K$.

Assume that the symbols $s_p$ and $s_u$ are in the same group (hence they are not orthogonal), while the symbols $s_q$ and $s_v$ are in another group (hence they are orthogonal to $s_p$ and $s_u$), we write $\{p, u\} \subset \mathcal{G}(p)$ and $\{q, v\} \not\subset \mathcal{G}(p)$ where $\mathcal{G}(p)$ represents a set of symbol indices that are in the same group as $s_p$, including $s_p$; similarly, $\{q, v\} \subset \mathcal{G}(q)$ and $\{p, u\} \not\subset \mathcal{G}(q)$. In order to achieve orthogonality among the symbols of different groups, e.g. between symbols $s_p$ and $s_q$, the summation terms included in the boxes in (7) are required to be zero. A way to achieve this is to make $\mathcal{A}_p^T\mathcal{A}_q$ and $\mathcal{A}_q^T\mathcal{A}_p$ (likewise $\mathcal{A}_p^T\mathcal{A}_v$, $\mathcal{A}_u^T\mathcal{A}_q$, $\mathcal{A}_u^T\mathcal{A}_v$ etc.) skew-symmetric, due to *Lemma 1* as stated below.

*Lemma 1*: If a matrix $\mathbf{M}$ of size $v \times v$ is skew-symmetric (i.e. $\mathbf{M}^T = -\mathbf{M}$), then $\mathbf{v}^T \mathbf{M} \mathbf{v} = 0$ for any vector $\mathbf{v}$ of size $v \times 1$.

Proof of *Lemma 1*: Let $c = \mathbf{v}^T \mathbf{M} \mathbf{v}$. Since $c$ is a scalar, $c^T = c$. If $\mathbf{M}^T = -\mathbf{M}$, then $c + c^T = \mathbf{v}^T \mathbf{M} \mathbf{v} + \mathbf{v}^T \mathbf{M}^T \mathbf{v} = 0$. Hence $c = 0$ if $\mathbf{M}$ is skew-symmetric, and *Lemma 1* is proved. ∎

*Theorem 1*: By ensuring that the dispersion matrices $\mathbf{A}_p$ and $\mathbf{A}_q$ of symbols $s_p$ and $s_q$ respectively meet the *Quasi-Orthogonality (QO) Constraint* specified in (8), their corresponding $\mathcal{A}_p^T \mathcal{A}_q$ and $\mathcal{A}_q^T \mathcal{A}_p$ will be skew-symmetric, and $s_p$ will be orthogonal to $s_q$.

$$\mathbf{A}_p^H \mathbf{A}_q = -\mathbf{A}_q^H \mathbf{A}_p \qquad \text{for } 1 \leq p, q \leq 2K \text{ and } q \notin \mathcal{G}(p) \qquad (8)$$

Proof of *Theorem 1*:

From (8),

$$\begin{aligned}
&\mathbf{A}_p^H \mathbf{A}_q = -\mathbf{A}_q^H \mathbf{A}_p \\
&\Rightarrow (\mathbf{A}_p^R + j\mathbf{A}_p^I)^H (\mathbf{A}_q^R + j\mathbf{A}_q^I) = -(\mathbf{A}_q^R + j\mathbf{A}_q^I)^H (\mathbf{A}_p^R + j\mathbf{A}_p^I) \\
&\Rightarrow \begin{cases} \text{real part}: (\mathbf{A}_p^R)^T \mathbf{A}_q^R + (\mathbf{A}_p^I)^T \mathbf{A}_q^I = -(\mathbf{A}_q^R)^T \mathbf{A}_p^R - (\mathbf{A}_q^I)^T \mathbf{A}_p^I \\ \text{imag part}: (\mathbf{A}_p^R)^T \mathbf{A}_q^I - (\mathbf{A}_p^I)^T \mathbf{A}_q^R = (\mathbf{A}_q^I)^T \mathbf{A}_p^R - (\mathbf{A}_q^R)^T \mathbf{A}_p^I \end{cases}
\end{aligned} \qquad (9)$$

Define

$$\begin{aligned}
\mathbf{M} &\triangleq \mathrm{Re}(\mathbf{A}_p^H \mathbf{A}_q) = (\mathbf{A}_p^R)^T \mathbf{A}_q^R + (\mathbf{A}_p^I)^T \mathbf{A}_q^I \\
\mathbf{N} &\triangleq \mathrm{Im}(\mathbf{A}_p^H \mathbf{A}_q) = (\mathbf{A}_p^R)^T \mathbf{A}_q^I - (\mathbf{A}_p^I)^T \mathbf{A}_q^R
\end{aligned}.$$

Then we know from (9) that $\mathbf{M}$ is skew-symmetric and $\mathbf{N}$ is symmetric. As a result, $\mathcal{A}_p^T \mathcal{A}_q = \begin{bmatrix} \mathbf{M} & -\mathbf{N} \\ \mathbf{N} & \mathbf{M} \end{bmatrix}$ is skew-symmetric. Similar conclusion can be drawn on $\mathcal{A}_q^T \mathcal{A}_p$. By *Lemma 1*, we then know that $\mathbf{H}^T \mathbf{H}$ in (7) is always block-diagonal. Hence *Theorem 1* is proved. ∎

It can be shown that all the QO-STBCs proposed in the literature, such as [2-4], follow the algebraic structure specified in (8). With this algebraic structure, we shall now examine the effect of CR on the decoding complexity of a QO-STBC.

## B. *Group Structure of QO-STBCs with CR*

We now examine the eight dispersion matrices of the code Q4 (listed in Appendix A as matrices $\mathbf{A}_1$, $\mathbf{A}_2$, …, $\mathbf{A}_8$) according to the derived QO-Constraint in (8). The fulfillment of QO-Constraint of $\mathbf{A}_1$, $\mathbf{A}_2$, …, $\mathbf{A}_8$ is shown in Table 1(a). For example, Table 1(a) shows that $\mathbf{A}_1$ is orthogonal to all the other dispersion matrices except $\mathbf{A}_4$. Likewise, each of $\mathbf{A}_2$, …, $\mathbf{A}_8$ are orthogonal to all but one of the other dispersion matrices. By re-arranging the rows and columns of Table 1(a) to obtain Table 1(b), it is clear that, the dispersion matrices $\mathbf{A}_1$, $\mathbf{A}_2$, …, $\mathbf{A}_8$ of the code Q4 can be grouped into four orthogonal groups, $\{\mathbf{A}_1, \mathbf{A}_4\}$; $\{\mathbf{A}_2, \mathbf{A}_3\}$; $\{\mathbf{A}_5, \mathbf{A}_8\}$; $\{\mathbf{A}_6, \mathbf{A}_7\}$, as depicted in Figure 1(a). Since there are only two non-orthogonal dispersion matrices (modulating two real symbols) in each group, the ML decoding of Q4 can be achieved by joint detection of two real symbols, instead of two complex symbols as reported in [2].

Table 2(a) examines the fulfillment of QO-Constraint for the dispersion matrices of the constellation-rotated Q4, i.e. Q4_CR, (listed in Appendix B as matrices $\mathbf{A}_{CR\_1}$, $\mathbf{A}_{CR\_2}$, …, $\mathbf{A}_{CR\_8}$). By re-arranging the rows and columns of Table 2(a) to obtain Table 2(b), it is clear that the dispersion matrices of Q4_CR can be grouped into only two orthogonal groups, $\{\mathbf{A}_{CR\_1}, \mathbf{A}_{CR\_4}, \mathbf{A}_{CR\_5}, \mathbf{A}_{CR\_8}\}$; $\{\mathbf{A}_{CR\_2}, \mathbf{A}_{CR\_3}, \mathbf{A}_{CR\_6}, \mathbf{A}_{CR\_7}\}$, as depicted in Figure 1(b). Since there are four non-orthogonal dispersion matrices (modulating four real symbols) in each group, it implies that in order to achieve full diversity using CR, the ML decoder for Q4_CR needs to jointly decode four real symbols, rather than two real symbols before CR.

It can similarly be shown that the rate-3/4 QO-STBC for eight transmit antennas proposed in [2] requires joint detection of two real symbols before CR (denoted herein as the Q8 code), and four real symbols after

CR [8] (denoted herein as the Q8_CR code). Likewise the rate-1 QO-STBC for eight transmit antennas proposed in [3] requires joint detection of four real symbols before CR (denoted herein as the T8 code), and four real symbols after CR [10] (denoted herein as the T8_CR code). This is summarized in Table 3.

## IV. GROUP-CONSTRAINED LINEAR TRANSFORMATION (GCLT)

### A. Definition of GCLT

In order to optimize a QO-STBC to achieve full diversity and maximum coding gain, while maintaining the original symbol groupings and hence the decoding complexity, we propose the *Group-Constrained Linear Transformation* (GCLT) as defined in *Proposition 1*.

*Proposition 1*: By linearly combining the dispersion matrices **A** within a group in accordance with (10) and (11) we can obtain a new set of dispersion matrices $\mathbf{A_{LT}}$ that will satisfy the QO-Constraint with the same symbol grouping structure as the original **A** matrices. Hence the transformation rules (10) and (11) do not destroy the quasi-orthogonal structure, nor change the number of quasi-orthogonal groups, of a QO-STBC.

$$\tilde{\mathbf{A}}_{\mathbf{LT}\_q} = \sum_{v \in \mathcal{G}(q)} \alpha_{q,v} \mathbf{A}_v \qquad 1 \leq q \leq 2K \qquad (10)$$

$$\mathbf{A}_{\mathbf{LT}\_q} = c_q \tilde{\mathbf{A}}_{\mathbf{LT}\_q} \qquad 1 \leq q \leq 2K \qquad (11)$$

where $\alpha_{q,v}$ is the *GCLT parameters* and are real constants. The scalar factor $c_q = \sqrt{\dfrac{TN_t}{K} \Big/ \operatorname{tr}(\tilde{\mathbf{A}}_{\mathbf{LT}\_q}^{\mathrm{H}} \tilde{\mathbf{A}}_{\mathbf{LT}\_q})}$ is to ensure that the dispersion matrices of the QO-STBC, after GCLT, satisfy the power distribution constraint in (2).

Proof of *Proposition 1*:

Applying (8) with $\mathbf{A}_p \rightarrow \mathbf{A}_{\mathbf{LT}\_p}$ and $\mathbf{A}_q \rightarrow \mathbf{A}_{\mathbf{LT}\_q}$ gives

$$\begin{aligned}
& \mathbf{A}_{LT\_p}^{H} \mathbf{A}_{LT\_q} + \mathbf{A}_{LT\_q}^{H} \mathbf{A}_{LT\_p} \qquad 1 \leq p, q \leq 2K \text{ and } p \notin \mathcal{G}(q) \\
& = c_p \big(\sum_{u \in \mathcal{G}(p)} \alpha_{p,u} \mathbf{A}_u\big)^H c_q \big(\sum_{v \in \mathcal{G}(q)} \alpha_{q,v} \mathbf{A}_v\big) + c_q \big(\sum_{v \in \mathcal{G}(q)} \alpha_{q,v} \mathbf{A}_v\big)^H c_p \big(\sum_{u \in \mathcal{G}(p)} \alpha_{p,u} \mathbf{A}_u\big) \\
& = c_p c_q \big(\sum_{u \in \mathcal{G}(p)} \sum_{v \in \mathcal{G}(q)} \alpha_{p,u} \alpha_{q,v} \mathbf{A}_u^H \mathbf{A}_v\big) + c_p c_q \big(\sum_{v \in \mathcal{G}(q)} \sum_{u \in \mathcal{G}(p)} \alpha_{q,v} \alpha_{p,u} \mathbf{A}_v^H \mathbf{A}_u\big) \\
& = c_p c_q \sum_{u \in \mathcal{G}(p)} \sum_{v \in \mathcal{G}(q)} \alpha_{p,u} \alpha_{q,v} \underbrace{(\mathbf{A}_u^H \mathbf{A}_v + \mathbf{A}_v^H \mathbf{A}_u)}_{=0 \text{ as per QO Constraint}} \\
& = 0
\end{aligned} \qquad (12)$$

Since the above expression is equal to zero, matrices {$\mathbf{A}_{LT}$} satisfy QO-Constraint (8) as matrices {$\mathbf{A}$} do, hence *Proposition 2* is proven. ∎

## B. Optimization of GCLT Parameters

The GCLT parameters, $\alpha$, can be chosen such that certain performance criteria, such as the rank and determinant criteria in [12], optimize the resultant {$\mathbf{A}_{LT}$}. To provide a systematic way to optimize the GCLT parameters in (10), Multi-dimensional Lattice Rotation (MLR) technique in [13] can be employed. For simplicity, consider a QO-STBC with two real symbols per group (such as code Q4). Assume that the matrices $\mathbf{A}_q$ and $\mathbf{A}_v$ are in the same group, i.e. $\{q,v\} = \mathcal{G}(q) = \mathcal{G}(v)$, the GCLT of the dispersion matrices can be expressed as follows:

$$\begin{bmatrix} \tilde{\mathbf{A}}_{LT\_q} \\ \tilde{\mathbf{A}}_{LT\_v} \end{bmatrix} = \left(\mathbf{I}_{T \times T} \otimes \begin{bmatrix} \alpha_{q,q} & \alpha_{q,v} \\ \alpha_{v,q} & \alpha_{v,v} \end{bmatrix}\right) \begin{bmatrix} \mathbf{A}_q \\ \mathbf{A}_v \end{bmatrix} = \left(\mathbf{I}_{T \times T} \otimes \mathbf{L}_{MLR}\right) \begin{bmatrix} \mathbf{A}_q \\ \mathbf{A}_v \end{bmatrix} \qquad (13)$$

where $\otimes$ represents the Kronecker product, and $\mathbf{I}_{T \times T}$ is an identity matrix of size $T \times T$, and $\mathbf{L}_{MLR}$ is an orthogonal matrix as specified in [13]. For a two-dimensional case, $\mathbf{L}_{MLR}$ maps four GCLT parameters into one variable $\theta$ using:

$$\mathbf{L}_{MLR} = \begin{bmatrix} \cos(\theta) & \sin(\theta) \\ -\sin(\theta) & \cos(\theta) \end{bmatrix} \qquad (14)$$

Hence, $\alpha_{q,q} = \alpha_{v,v} = \cos(\theta); \alpha_{q,v} = -\alpha_{v,q} = \sin(\theta)$ in this case. This facilitates the search or analysis of the optimum GCLT parameters. Denoting the Q4 code after GCLT as Q4_LT, we provide here an analytical derivation for the optimization of its GCLT parameters. First, the determinant expression for the codeword distance matrix of Q4_LT is derived as follows:

$$\det = \begin{bmatrix} ((\tilde{\Delta}_1+\tilde{\Delta}_4)^2 + (\tilde{\Delta}_2-\tilde{\Delta}_3)^2 + (\tilde{\Delta}_5+\tilde{\Delta}_8)^2 + (\tilde{\Delta}_6-\tilde{\Delta}_7)^2) \times \\ ((\tilde{\Delta}_1-\tilde{\Delta}_4)^2 + (\tilde{\Delta}_2+\tilde{\Delta}_3)^2 + (\tilde{\Delta}_5-\tilde{\Delta}_8)^2 + (\tilde{\Delta}_6+\tilde{\Delta}_7)^2) \end{bmatrix}^2 \quad (15)$$

where $\tilde{\Delta}_q = \Delta_q \cos\theta - \Delta_v \sin\theta$ and $\tilde{\Delta}_v = \Delta_q \sin\theta + \Delta_v \cos\theta$ for $\{(q,v)\} \in \{(1,4),(2,3),(5,8),(6,7)\}$, and $\Delta_q$ represents the possible error in the real PAM symbol $s_q$ (remembering that a QAM symbol $x_q$ is expressed in terms of two real PAM symbols, i.e. $x_q = s_q + js_{K+q}$ where $K$ is the number of complex symbols being transmitted in a STBC codeword).

Since the symbol grouping of Q4_LT is such that $s_1$ and $s_4$ are in a group, and they are independent of (i.e. orthogonal to) the other symbols in the ML decoding operation, without loss of generality, it can be assumed that only $s_1$ and $s_4$ have errors and the other symbols are error-free [6]. As a result, the worst-case (i.e. minimum) determinant value in (15) can be simplified to:

$$\begin{aligned}
\det &= \left[(\tilde{\Delta}_1+\tilde{\Delta}_4)^2 \times (\tilde{\Delta}_1-\tilde{\Delta}_4)^2\right]^2 \quad \text{assuming that } \tilde{\Delta}_2=\tilde{\Delta}_3=\tilde{\Delta}_5=\tilde{\Delta}_6=\tilde{\Delta}_7=\tilde{\Delta}_8=0 \\
&= \left[(\tilde{\Delta}_1)^2 - (\tilde{\Delta}_4)^2\right]^4 \\
&= \left[(\Delta_1\cos\theta - \Delta_4\sin\theta)^2 - (\Delta_1\sin\theta + \Delta_4\cos\theta)^2\right]^4 \\
&= \left[\Delta_1^2\cos(2\theta) - 2\Delta_1\Delta_4\sin(2\theta) - \Delta_4^2\cos(2\theta)\right]^4
\end{aligned} \quad (16)$$

*4-QAM Constellation*

Consider first 4-QAM constellation. The I and Q components of a 4-QAM symbol can be viewed as two independent 2-PAM symbols. Hence $\Delta_1, \Delta_4 \in \{0, \pm d_{min}\}$ where $d_{min}$ is the minimum Euclidean distance between two constellation points as shown in Figure 2, and $\Delta_1$ and $\Delta_4$ cannot be both zero in (16). To maximize the minimum determinant value in (16) based on the rank and determinant criteria in [12], the following four cases of $(\Delta_1, \Delta_4)$ and their resultant determinant values as per (16) are considered:

$$\text{Case 1: } (\Delta_1, \Delta_4) = \pm (0, d_{min}) \rightarrow \det_1 = d_{min}^8 \left[\cos(2\theta)\right]^4 \quad (17)$$

$$\text{Case 2: } (\Delta_1, \Delta_4) = \pm (d_{min}, 0) \rightarrow \det_2 = d_{min}^8 \left[\cos(2\theta)\right]^4 \quad (18)$$

$$\text{Case 3: } (\Delta_1, \Delta_4) = \pm (d_{min}, d_{min}) \rightarrow \det_3 = d_{min}^8 \left[-2\sin(2\theta)\right]^4 \quad (19)$$

$$\text{Case 4: } (\Delta_1, \Delta_4) = \pm (d_{min}, -d_{min}) \rightarrow \det_4 = d_{min}^8 \left[2\sin(2\theta)\right]^4 \quad (20)$$

Note that $\det_1 = \det_2$ and $\det_3 = \det_4$. In order to maximize the smaller value between $\det_1$ and $\det_3$, we equate $\det_1$ and $\det_3$ to get:

$$\begin{aligned} &\cos(2\theta_{opt}) = 2\sin(2\theta_{opt}) \\ &\Rightarrow \tan(2\theta_{opt}) = 1/2 \\ &\Rightarrow \theta_{opt} = \frac{1}{2}\tan^{-1}(\frac{1}{2}) = 13.28^0 \end{aligned} \quad (21)$$

So the optimum GCLT parameters for Q4_LT are:

$$\begin{aligned} \alpha_{q,q} &= \alpha_{v,v} = \cos(\theta_{opt}) = \cos(13.28^0) \\ \alpha_{q,v} &= -\alpha_{v,q} = \sin(\theta_{opt}) = \sin(13.28^0) \end{aligned} \quad (22)$$

where $\{(q,v)\} \in \{(1,4),(2,3),(5,8),(6,7)\}$, and the minimum determinant value of the codeword distance matrix is $0.64 d_{min}^8$. Compared with Q4_CR, which has a minimum determinant value $d_{min}^8$ [10], Q4_LT has a slightly smaller minimum determinant value (which will be shown later to give less than 0.5dB loss in coding gain), but the ML decoding of Q4_LT requires the joint detection of half the number of symbols as required by Q4_CR.

*M-ary QAM Constellation*

We now derive the optimum GCLT parameters of Q4_LT for larger square QAM constellations. Consider the *M*-ary square-QAM constellation, where the I and Q components of a symbol can be viewed as two independent $\sqrt{M}$-ary PAM symbols. The following four cases of $(\Delta_1, \Delta_4)$ and their resultant determinant values as per (16) are considered:

Case 1: $(\Delta_1, \Delta_4) = \pm (0, nd_{min})$ → $\det_5 = d_{min}^8 \left[ n^2 \cos(2\theta) \right]^4$ (23)

Case 2: $(\Delta_1, \Delta_4) = \pm (md_{min}, 0)$ → $\det_6 = d_{min}^8 \left[ m^2 \cos(2\theta) \right]^4$ (24)

Case 3: $(\Delta_1, \Delta_4) = \pm (md_{min}, nd_{min})$ → $\det_7 = d_{min}^8 \left[ (m^2 - n^2)\cos(2\theta) - 2mn\sin(2\theta) \right]^4$ (25)

Case 4: $(\Delta_1, \Delta_4) = \pm (md_{min}, -nd_{min})$ → $\det_8 = d_{min}^8 \left[ (m^2 - n^2)\cos(2\theta) + 2mn\sin(2\theta) \right]^4$ (26)

where $d_{min}$ represents the minimum Euclidean distance between the PAM constellation points, $m$ and $n$ are integers such that $1 \leq m, n \leq \sqrt{M} - 1$, and $M$ is the cardinality of the QAM constellation.

To maximize the smaller value of $\det_5$ to $\det_8$ for all valid values of $m$ and $n$, consider first the smallest value of $m = n = 1$. For this case, $\det_5$ to $\det_8$ are identical to $\det_1$ to $\det_4$, hence the optimum $\theta$ value for (23) to (26) is the same as that for (21), i.e. $\theta_{opt} = \tfrac{1}{2} \tan^{-1}(\tfrac{1}{2})$, and the corresponding $\det_5$ to $\det_8$ values are identical and equal to $0.64 d_{min}^8$.

Next, consider $m, n > 1$. In this case, it can be shown that with $\theta = \tfrac{1}{2} \tan^{-1}(\tfrac{1}{2})$,

$$\det_5 = n^8 \left( 0.64 d_{min}^8 \right)$$

$$\det_6 = m^8 \left( 0.64 d_{min}^8 \right)$$

$$\det_7 = \left[ (m^2 - mn - n^2) \right]^4 \left( 0.64 d_{min}^8 \right)$$

$$\det_8 = \left[ (m^2 + mn - n^2) \right]^4 \left( 0.64 d_{min}^8 \right)$$

which are all greater than or equal to $0.64 d_{min}^8$ for $m, n > 1$. Hence the worst-case (i.e. minimum) determinant value for Q4_LT with $M$-ary square-QAM constellation occurs when $m = n = 1$, and it is optimized when $\theta = \theta_{opt} = \tfrac{1}{2} \tan^{-1}(\tfrac{1}{2})$. Therefore, we have shown that the optimum GCLT parameters derived in (22) apply to all QAM size.

To show the above result graphically, the determinant values of Q4_LT with square 16-QAM constellation is plotted as a function of $\theta$ in Figure 3. For the square 16-QAM, the values of $m$ and $n$ can each take values of 1, 2 or 3. A few combinations of $m$ and $n$ are shown in Figure 3 as illustration. We can see that at $\theta_{opt}$, all the determinant values corresponding to all possible values of $m$ and $n$ are greater than or equal to the determinant values corresponding to the case of $m=n=1$, and the optimum determinant values corresponding to $m=n=1$ occurs at $\theta_{opt}$.

Denoted as $\mathbf{A}_{LT\_1}$, $\mathbf{A}_{LT\_2}$, ..., $\mathbf{A}_{LT\_8}$, the dispersion matrices of Q4_LT obtained based on the optimum GCLT parameters derived in (22) are shown in Appendix C. As shown in Figure 1(c), Q4_LT has exactly the same symbol grouping structure as Q4.

It can similarly be shown that the optimum GCLT parameters for Q8 with M-ary QAM constellation are:

$$\begin{aligned} \alpha_{q,q} &= \alpha_{v,v} = \cos(13.28^0) \\ \alpha_{q,v} &= -\alpha_{v,q} = \sin(13.28^0) \end{aligned} \quad (27)$$

where $\{(q,v)\} \in \{(1,10),(2,11),(3,12),(4,7),(5,8),(6,9)\}$. The resultant code, denoted as Q8_LT, has minimum determinant value of $0.4096(\sqrt{4/3}d_{min})^{16}$, as compared with $(\sqrt{4/3}d_{min})^{16}$ for Q8_CR.

In [13], the rate-1 QO-STBC for eight transmit antennas T8 requires a joint detection of four real symbols. Its $\mathbf{L}_{MLR}$ matrix corresponding to (13):

$$\mathbf{L}_{MLR} = \prod_{1 \leq i \leq 3,\ i+1 \leq k \leq 4} \mathbf{G}(i,k,\theta_{ik}) \quad (28)$$

where $\mathbf{G}(i, k, \theta_{ik})$ is a 4×4 matrix with entries at $(i, i)$ and $(k, k)$ equal to $\cos(\theta_{ik})$, entry at $(i, k)$ equals to $\sin(\theta_{ik})$, and entry at $(k, i)$ equals to $-\sin(\theta_{ik})$, one on the remaining diagonal positions and zero elsewhere.

$\mathbf{G}(i, k, \theta_{ik})$ basically models a counter-clockwise rotation by $\theta$ degree with respect to the $(i, k)$ plane. For example, for $i = 2, k = 3$, the **G** matrix becomes:

$$\mathbf{G}(2,3,\theta_{23}) = \begin{bmatrix} 1 & 0 & 0 & 0 \\ 0 & \cos(\theta_{23}) & \sin(\theta_{23}) & 0 \\ 0 & -\sin(\theta_{23}) & \cos(\theta_{23}) & 0 \\ 0 & 0 & 0 & 1 \end{bmatrix} \tag{29}$$

The GCLT parameters are related to $\mathbf{L}_{MLR}$ by the following relationship:

$$\begin{bmatrix} \alpha_{p,p} & \alpha_{p,q} & \alpha_{p,m} & \alpha_{p,n} \\ \alpha_{q,p} & \alpha_{q,q} & \alpha_{q,m} & \alpha_{q,n} \\ \alpha_{m,p} & \alpha_{m,q} & \alpha_{m,m} & \alpha_{m,n} \\ \alpha_{n,p} & \alpha_{n,q} & \alpha_{n,m} & \alpha_{n,n} \end{bmatrix} = \mathbf{L}_{MLR} \tag{30}$$

where $\{(p, q, m, n)\} \in \{(1, 4, 6, 7), (2, 3, 5, 8), (9, 12, 14, 15), (10, 11, 13, 16)\}$ for T8. The mathematical derivation of optimum GCLT parameters for T8 is difficult because of the large dimension involved. Hence we rely on computer search to find the optimization solution. The best solution found is: $\theta_{12} = -45.66^0$, $\theta_{23} = 9.43^0$, $\theta_{34} = -46.11^0$, $\theta_{14} = 37.78^0$, $\theta_{13} = 9.13^0$, $\theta_{24} = 44.24^0$.

The diversity product $\zeta$, defined in (31), is a good indicator of the decoding performance of a STBC [8].

$$\zeta = \frac{1}{2\sqrt{N_t}} |\text{Det}_{\min}|^{1/(2T)} \tag{31}$$

The diversity product of Q4_CR, Q4_LT, Q8_CR, Q8_LT, T8_CR and T8_LT with 4-QAM are listed in Table 3. We can see that Q4_LT, Q8_LT and T8_LT achieve a lower diversity product (hence coding gain) than Q4_CR, Q8_CR and T8_CR, but Q4_LT, Q8_LT and T8_LT only need to jointly decode half of the symbols as required by Q4_CR, Q8_CR and T8_CR respectively. Furthermore, it will be shown later on that, despite their reduced coding gains, Q4_LT, Q8_LT and T8_LT have negligible performance loss compared to Q4_CR, Q8_CR and T8_CR respectively.

The reduction in diversity product (and hence coding gain) of the QO-STBC with GCLT over QO-STBC with CR can be explained as follows: since GCLT only restrict the symbols in a group to be linearly transformed, while CR combines symbols across different group, hence CR has a higher degree of freedom when performing the minimum determinant optimization, and hence it achieves a higher diversity product. However, it should be noted that the higher diversity product achieved by CR comes at the expense of an increased decoding complexity.

### C. ML Decoding

The ML decoding metrics of Q4_LT are shown in (32).

$$
\begin{aligned}
f_1(s_1,s_4) &= \sum_{r=1}^{N_r}\left[(\sum_{n=1}^{4}|h_{n,r}|^2)(|as_1-bs_4|^2+|bs_1+as_4|^2)+2\,\mathrm{Re}\{(as_1-bs_4)\alpha+(bs_1+as_4)\beta+(as_1-bs_4)(bs_1+as_4)\gamma\}\right] \\
f_2(s_2,s_3) &= \sum_{r=1}^{N_r}\left[(\sum_{n=1}^{4}|h_{n,r}|^2)(|as_2-bs_3|^2+|bs_2+as_3|^2)+2\,\mathrm{Re}\{(as_2-bs_3)\chi+(bs_2+as_3)\delta+(as_2-bs_3)(bs_2+as_3)\varphi\}\right] \\
f_3(s_5,s_8) &= \sum_{r=1}^{N_r}\left[(\sum_{n=1}^{4}|h_{n,r}|^2)(|as_5-bs_8|^2+|bs_5+as_8|^2)+2\,\mathrm{Re}\{j(as_5-bs_8)\alpha+j(bs_5+as_8)\beta+(as_5-bs_8)(bs_5+as_8)\gamma\}\right] \\
f_4(s_6,s_7) &= \sum_{r=1}^{N_r}\left[(\sum_{n=1}^{4}|h_{n,r}|^2)(|as_6-bs_7|^2+|bs_6+as_7|^2)+2\,\mathrm{Re}\{j(as_6-bs_7)\chi+j(bs_6+as_7)\delta+(as_6-bs_7)(bs_6+as_7)\varphi\}\right]
\end{aligned}
\quad (32)
$$

where $\alpha=-h_{1,r}r_1^*-h_{2,r}^*r_2-h_{3,r}^*r_3-h_{4,r}r_4^*$, $\beta=-h_{4,r}r_1^*+h_{3,r}^*r_2+h_{2,r}^*r_3-h_{1,r}r_4^*$, $\gamma=2\,\mathrm{Re}(h_{1,r}h_{4,r}^*-h_{2,r}h_{3,r}^*)$,

$\chi=-h_{2,r}r_1^*+h_{1,r}^*r_2-h_{4,r}^*r_3+h_{3,r}r_4^*$, $\delta=-h_{3,r}r_1^*-h_{4,r}^*r_2+h_{1,r}^*r_3+h_{2,r}r_4^*$, $\varphi=2\,\mathrm{Re}(-h_{1,r}h_{4,r}^*+h_{2,r}h_{3,r}^*)$,

a = cos(13.28$^0$), b = sin(13.28$^0$), and each $s_i$ is a *real* symbol.

Each of the decoding metrics shown above depends only on two *real* symbols, in contrast to (6) which shows that each decoding metric of Q4_CR relies on two *complex* symbols. Hence the proposed GCLT scheme can achieve a significant reduction of decoding complexity compared with the constellation rotation scheme. This complexity reduction is all the more significant for the larger QAM constellation size.

Although only Q4, Q8 from [2] and T8 from [3] are used as examples in this paper, the approach described in this paper can be used to achieve the same reduction in decoding complexity (i.e. halving of the number of symbols required for ML joint detection) for the other QO-STBCs reported in the literature [5-8,10] too.

*D. Decoding Performance*

The BER performance of the Q4, Q4_CR and Q4_LT codes are compared in this section, using the O-STBC from [1] as performance benchmark. Since Q4, Q4_CR and Q4_LT are full-rate codes, while the O-STBC G4C is a half-rate code, 16-QAM constellation is used for the latter while 4-QAM constellation is used for the QO-STBCs in order to achieve the same spectral efficiency of 2 bits/s/Hz for all codes. In Figure 4, it is observed that both Q4_CR [8] and Q4_LT (constructed in this paper) achieve full transmit diversity as they have the same BER slope as the G4C. Q4_CR and Q4_LT also have lower BER than the G4C because they are full-rate codes with smaller QAM dimension and hence larger Euclidean distance. Although Q4_CR has slightly better performance (due to a larger minimum determinant value as shown in Table 3) than Q4_LT, their performance difference is less than 0.5dB. Q4_LT, however, needs joint detection of only two *real* symbols; hence it has a significantly lower decoding complexity than Q4_CR, which requires the joint detection of two *complex* symbols. Similar observations hold for the case with two receive antennas.

The comparisons between Q8_CR and Q8_LT, T8_CR and T8_LT with 4-QAM and one receive antennas is shown in Figure 5. It can be seen again that QO-STBC optimized with GCLT has only less than 0.5dB loss in decoding performance, but the number of symbols required for joint detection is halved as shown in Table 3.

It should be noted that we adopt the STBC signal model in [11] in which the dispersion matrices are used to modulate the *real and imaginary parts of a complex symbols*, as opposed to the STBC signal model in [5] in which the dispersion matrices are used to modulate the *complex symbols and their corresponding conjugate symbols*. As a result, we are able to get an insight into the decoding complexity of QO-STBC with GCLT versus QO-STBC with CR [5].

## V.  Conclusion

In this paper, we first derive the generic algebraic structure of QO-STBC, called Quasi-Orthogonality (QO) Constraint. It can be shown that all existing QO-STBCs are unified under this algebraic structure. Based on the derived QO Constraint, we find that the constellation rotation (CR) technique, which is commonly used to improve the decoding performance of a QO-STBC, actually increases the decoding complexity of the resultant QO-STBC, as the number of symbols required for joint detection in ML decoding is doubled after CR is applied. Hence we propose Group-Constrained Linear Transformation (GCLT) as a means to improve the decoding performance of a QO-STBC with QAM constellation without increasing the number of symbols required for joint detection. The optimum GCLT parameters for achieving maximum diversity and coding gains are derived analytically for square QAM constellations. Simulation results show that QO-STBC with GCLT can achieve full diversity at less than 0.5 dB loss in coding gain compared to QO-STBC with CR.

APPENDIX A

Dispersion Matrices of Q4 [2]:

$$\mathbf{A}_1 = \begin{bmatrix} 1 & 0 & 0 & 0 \\ 0 & 1 & 0 & 0 \\ 0 & 0 & 1 & 0 \\ 0 & 0 & 0 & 1 \end{bmatrix}, \mathbf{A}_2 = \begin{bmatrix} 0 & 1 & 0 & 0 \\ -1 & 0 & 0 & 0 \\ 0 & 0 & 0 & 1 \\ 0 & 0 & -1 & 0 \end{bmatrix}, \mathbf{A}_3 = \begin{bmatrix} 0 & 0 & 1 & 0 \\ 0 & 0 & 0 & 1 \\ -1 & 0 & 0 & 0 \\ 0 & -1 & 0 & 0 \end{bmatrix}, \mathbf{A}_4 = \begin{bmatrix} 0 & 0 & 0 & 1 \\ 0 & 0 & -1 & 0 \\ 0 & -1 & 0 & 0 \\ 1 & 0 & 0 & 0 \end{bmatrix},$$

$$\mathbf{A}_5 = \begin{bmatrix} j & 0 & 0 & 0 \\ 0 & -j & 0 & 0 \\ 0 & 0 & -j & 0 \\ 0 & 0 & 0 & j \end{bmatrix}, \mathbf{A}_6 = \begin{bmatrix} 0 & j & 0 & 0 \\ j & 0 & 0 & 0 \\ 0 & 0 & 0 & -j \\ 0 & 0 & -j & 0 \end{bmatrix}, \mathbf{A}_7 = \begin{bmatrix} 0 & 0 & j & 0 \\ 0 & 0 & 0 & -j \\ j & 0 & 0 & 0 \\ 0 & -j & 0 & 0 \end{bmatrix}, \mathbf{A}_8 = \begin{bmatrix} 0 & 0 & 0 & j \\ 0 & 0 & j & 0 \\ 0 & j & 0 & 0 \\ j & 0 & 0 & 0 \end{bmatrix}.$$

APPENDIX B

Dispersion Matrices of Q4_CR [8]:

$$\mathbf{A}_{CR\_1} = \begin{bmatrix} 1 & 0 & 0 & 0 \\ 0 & 1 & 0 & 0 \\ 0 & 0 & 1 & 0 \\ 0 & 0 & 0 & 1 \end{bmatrix}, \mathbf{A}_{CR\_2} = \begin{bmatrix} 0 & 1 & 0 & 0 \\ -1 & 0 & 0 & 0 \\ 0 & 0 & 0 & 1 \\ 0 & 0 & -1 & 0 \end{bmatrix},$$

$$\mathbf{A}_{CR\_3} = \frac{1}{\sqrt{2}} \begin{bmatrix} 0 & 0 & 1+j & 0 \\ 0 & 0 & 0 & 1-j \\ -1+j & 0 & 0 & 0 \\ 0 & -1-j & 0 & 0 \end{bmatrix}, \mathbf{A}_{CR\_4} = \frac{1}{\sqrt{2}} \begin{bmatrix} 0 & 0 & 0 & 1+j \\ 0 & 0 & -1+j & 0 \\ 0 & -1+j & 0 & 0 \\ 1+j & 0 & 0 & 0 \end{bmatrix},$$

$$\mathbf{A}_{CR\_5} = \begin{bmatrix} j & 0 & 0 & 0 \\ 0 & -j & 0 & 0 \\ 0 & 0 & -j & 0 \\ 0 & 0 & 0 & j \end{bmatrix}, \mathbf{A}_{CR\_6} = \begin{bmatrix} 0 & j & 0 & 0 \\ j & 0 & 0 & 0 \\ 0 & 0 & 0 & -j \\ 0 & 0 & -j & 0 \end{bmatrix},$$

$$\mathbf{A}_{CR\_7} = \frac{j}{\sqrt{2}} \begin{bmatrix} 0 & 0 & 1+j & 0 \\ 0 & 0 & 0 & -1+j \\ 1-j & 0 & 0 & 0 \\ 0 & -1-j & 0 & 0 \end{bmatrix}, \mathbf{A}_{CR\_8} = \frac{j}{\sqrt{2}} \begin{bmatrix} 0 & 0 & 0 & 1+j \\ 0 & 0 & 1-j & 0 \\ 0 & 1-j & 0 & 0 \\ 1+j & 0 & 0 & 0 \end{bmatrix}.$$

APPENDIX C

Dispersion Matrices of Q4_LT:

$$\mathbf{A}_{LT\_1} = \begin{bmatrix} a & 0 & 0 & b \\ 0 & a & -b & 0 \\ 0 & -b & a & 0 \\ b & 0 & 0 & a \end{bmatrix}, \mathbf{A}_{LT\_2} = \begin{bmatrix} 0 & a & b & 0 \\ -a & 0 & 0 & b \\ -b & 0 & 0 & a \\ 0 & -b & -a & 0 \end{bmatrix},$$

$$\mathbf{A}_{LT\_3} = \begin{bmatrix} 0 & -b & a & 0 \\ b & 0 & 0 & a \\ -a & 0 & 0 & -b \\ 0 & -a & b & 0 \end{bmatrix}, \mathbf{A}_{LT\_4} = \begin{bmatrix} -b & 0 & 0 & a \\ 0 & -b & -a & 0 \\ 0 & -a & -b & 0 \\ a & 0 & 0 & -b \end{bmatrix},$$

$$\mathbf{A}_{LT\_5} = \begin{bmatrix} ja & 0 & 0 & jb \\ 0 & -ja & jb & 0 \\ 0 & jb & -ja & 0 \\ jb & 0 & 0 & ja \end{bmatrix}, \mathbf{A}_{LT\_6} = \begin{bmatrix} 0 & ja & jb & 0 \\ ja & 0 & 0 & -jb \\ jb & 0 & 0 & -ja \\ 0 & -jb & -ja & 0 \end{bmatrix},$$

$$\mathbf{A}_{LT\_7} = \begin{bmatrix} 0 & -jb & ja & 0 \\ -jb & 0 & 0 & -ja \\ ja & 0 & 0 & jb \\ 0 & -ja & jb & 0 \end{bmatrix}, \mathbf{A}_{LT\_8} = \begin{bmatrix} -jb & 0 & 0 & ja \\ 0 & jb & ja & 0 \\ 0 & ja & jb & 0 \\ ja & 0 & 0 & -jb \end{bmatrix}.$$

where $a = \cos(13.28^0)$ and $b = \sin(13.28^0)$.

LIST OF ILLUSTRATIONS

|  Q4  | Q4_CR | Q4_LT |
|---|---|---|
| Group 1  Group 2<br>$A_1$ $A_4$   $A_2$ $A_3$<br>$A_5$ $A_8$   $A_6$ $A_7$<br>Group 3  Group 4 | Group 1   Group 2<br>$A_{CR1}$ $A_{CR4}$   $A_{CR2}$ $A_{CR3}$<br>$A_{CR5}$ $A_{CR8}$   $A_{CR6}$ $A_{CR7}$ | Group 1  Group 2<br>$A_{LT1}$ $A_{LT4}$   $A_{LT2}$ $A_{LT3}$<br>$A_{LT5}$ $A_{LT8}$   $A_{LT6}$ $A_{LT7}$<br>Group 3  Group 4 |
| (a) | (b) | (c) |

Figure 1 Grouping structure of dispersion matrices of Q4 [2], Q4 with CR (Q4_CR) [8] and Q4 with GCLT (Q4_LT)

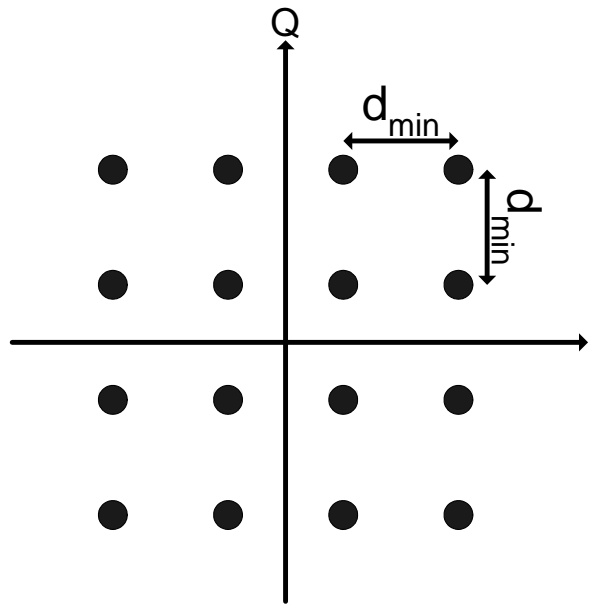

Figure 2 QAM-constellation and its minimum Euclidean distance $d_{min}$

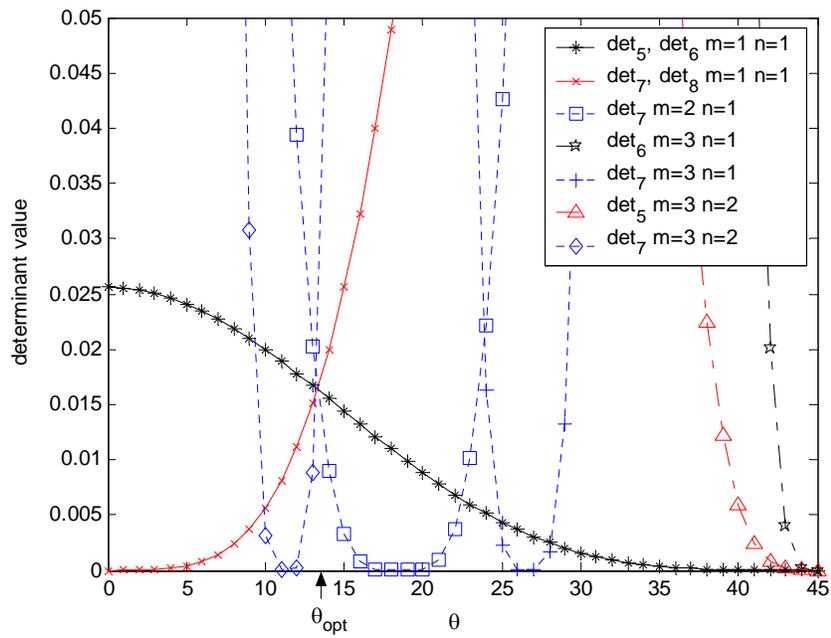

Figure 3 Determinant values versus angle of optimization

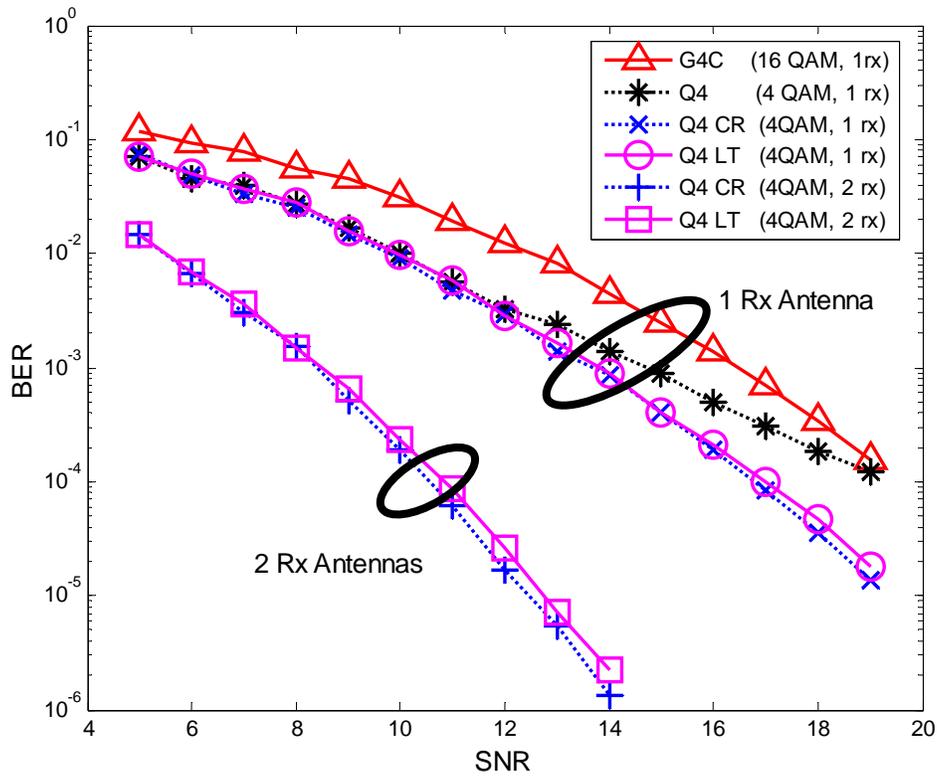

Figure 4 Simulation results of QO-STBCs for four transmit antennas with spectral efficiency of 2 bits/sec/Hz

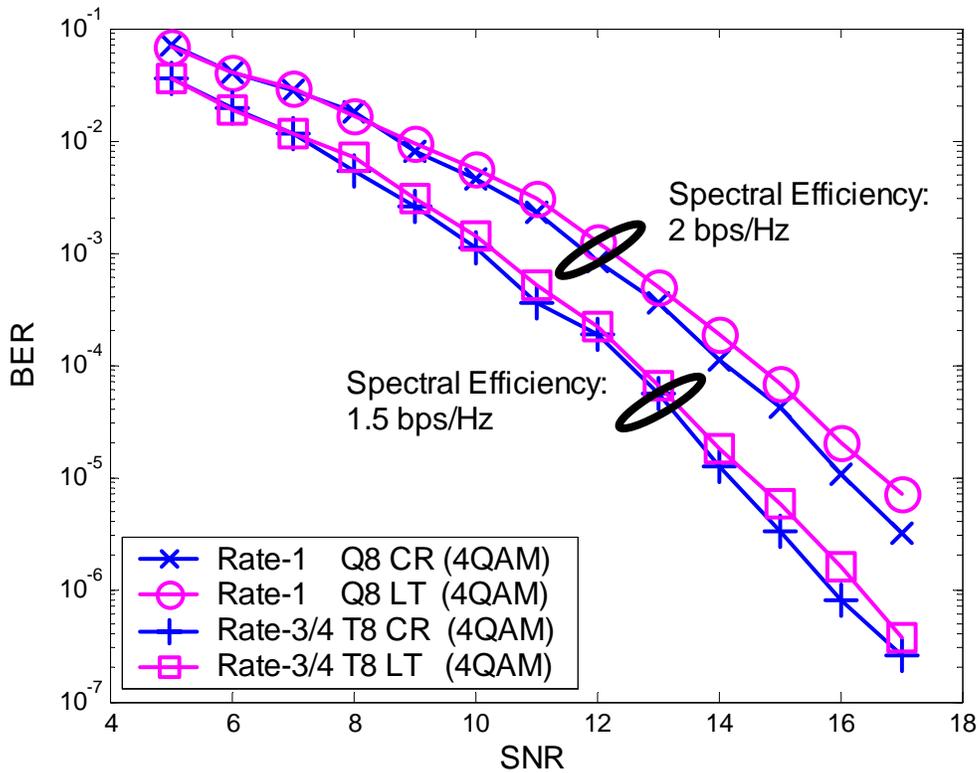

Figure 5 Simulation results of QO-STBCs for eight transmit and one receive antenna

Table 1 Non-fulfillment of Quasi-Orthogonality Constraint in (8) for the code Q4 [2]

(a) QO-Constraint fulfillment for Q4

|       | $A_1$ | $A_2$ | $A_3$ | $A_4$ | $A_5$ | $A_6$ | $A_7$ | $A_8$ |
|-------|-------|-------|-------|-------|-------|-------|-------|-------|
| $A_1$ | X     |       |       | X     |       |       |       |       |
| $A_2$ |       | X     | X     |       |       |       |       |       |
| $A_3$ |       | X     | X     |       |       |       |       |       |
| $A_4$ | X     |       |       | X     |       |       |       |       |
| $A_5$ |       |       |       |       | X     |       |       | X     |
| $A_6$ |       |       |       |       |       | X     | X     |       |
| $A_7$ |       |       |       |       |       | X     | X     |       |
| $A_8$ |       |       |       |       | X     |       |       | X     |

X : QO-Constraint is not fulfilled

(b) Re-arrange rows and columns of (a)

|       | $A_1$ | $A_4$ | $A_2$ | $A_3$ | $A_5$ | $A_8$ | $A_6$ | $A_7$ |
|-------|-------|-------|-------|-------|-------|-------|-------|-------|
| $A_1$ | X     | X     |       |       |       |       |       |       |
| $A_4$ | X     | X     |       |       |       |       |       |       |
| $A_2$ |       |       | X     | X     |       |       |       |       |
| $A_3$ |       |       | X     | X     |       |       |       |       |
| $A_5$ |       |       |       |       | X     | X     |       |       |
| $A_8$ |       |       |       |       | X     | X     |       |       |
| $A_6$ |       |       |       |       |       |       | X     | X     |
| $A_7$ |       |       |       |       |       |       | X     | X     |

X : QO-Constraint is not fulfilled

Table 2 Non-fulfillment of Quasi-Orthogonality Constraint in (8) for the code Q4_CR [8]

(a) QO-Constraint fulfillment for Q4_CR

|          | $A_{CR1}$ | $A_{CR2}$ | $A_{CR3}$ | $A_{CR4}$ | $A_{CR5}$ | $A_{CR6}$ | $A_{CR7}$ | $A_{CR8}$ |
|----------|-----------|-----------|-----------|-----------|-----------|-----------|-----------|-----------|
| $A_{CR1}$ | X         |           |           | X         |           |           |           |           |
| $A_{CR2}$ |           | X         | X         |           |           |           |           |           |
| $A_{CR3}$ |           | X         | X         |           |           | X         | X         |           |
| $A_{CR4}$ | X         |           |           | X         | X         |           |           | X         |
| $A_{CR5}$ |           |           |           | X         | X         |           |           | X         |
| $A_{CR6}$ |           |           | X         |           |           | X         | X         |           |
| $A_{CR7}$ |           |           | X         |           |           | X         | X         |           |
| $A_{CR8}$ |           |           |           | X         | X         |           |           | X         |

X : QO-Constraint is not fulfilled

(b) Rearrange rows and columns of (a)

|            | $A_{CR1}$ | $A_{CR4}$ | $A_{CR5}$ | $A_{CR8}$ | $A_{CR2}$ | $A_{CR3}$ | $A_{CR6}$ | $A_{CR7}$ |
|------------|-----------|-----------|-----------|-----------|-----------|-----------|-----------|-----------|
| $A_{CR1}$  | X         | X         |           |           |           |           |           |           |
| $A_{CR4}$  | X         | X         | X         | X         |           |           |           |           |
| $A_{CR5}$  |           | X         | X         | X         |           |           |           |           |
| $A_{CR8}$  |           | X         | X         | X         |           |           |           |           |
| $A_{CR2}$  |           |           |           |           | X         | X         |           |           |
| $A_{CR3}$  |           |           |           |           | X         | X         | X         | X         |
| $A_{CR6}$  |           |           |           |           |           | X         | X         | X         |
| $A_{CR7}$  |           |           |           |           |           | X         | X         | X         |

X : QO-Constraint is not fulfilled

Table 3 Comparison of QO-STBCs with CR and GCLT

|                                               |          | No. of real symbols for ML joint detection | Diversity product $\zeta$ for 4-QAM |
|-----------------------------------------------|----------|--------------------------------------------|-------------------------------------|
| Rate-1 QO-STBC for four transmit antennas     | Q4 [2]   | 2                                          | Non full diversity                  |
|                                               | Q4_CR [8]| 4                                          | 0.3536                              |
|                                               | **Q4_LT**| **2**                                      | **0.3344**                          |
| Rate-3/4 QO-STBC for eight transmit antennas  | Q8 [2]   | 2                                          | Non full diversity                  |
|                                               | Q8_CR [8]| 4                                          | 0.2887                              |
|                                               | **Q8_LT**| **2**                                      | **0.2730**                          |
| Rate-1 QO-STBC for eight transmit antennas    | T8 [3]   | 4                                          | Non full diversity                  |
|                                               | T8_CR [10]| 8                                         | 0.2187                              |
|                                               | **T8_LT**| **4**                                      | **0.1531**                          |